# Learning Program Component Order

Steven P. Reiss and Qi Xin
Brown University, Providence, RI
{spr,qx5}@cs.brown.edu

*Abstract*—Successful programs are written to be maintained. One aspect to this is that programmers order the components in the code files in a particular way. This is part of programming style. While the conventions for ordering are sometimes given as part of a style guideline, such guidelines are often incomplete and programmers tend to have their own more comprehensive orderings in mind. This paper defines a model for ordering program components and shows how this model can be learned from sample code. Such a model is a useful tool for a programming environment in that it can be used to find the proper location for inserting new components or for reordering files to better meet the needs of the programmer. The model is designed so that it can be fine-tuned by the programmer. The learning framework is evaluated both by looking at code with known style guidelines and by testing whether it inserts existing components into a file correctly.

*Keywords—Program style, component ordering, programming environments.*

## I. MOTIVATION

Programming style is key to program maintenance. Reading and understanding code for maintenance is significantly easier if the code follows a consistent style. Part of this style is the way the various program components, i.e. fields, functions, methods, classes, interfaces, etc., are ordered. Using a consistent program order can greatly simplify understanding code. For example, Google notes that "the order you choose for the members and initializers of your class can have a great effect on learnability" [10].

For these reasons, the order of program components is often included in the style guidelines that are developed by individuals, projects, and companies. While concentrating on local style, indentation, and naming conventions, these guidelines also specify how files should be organized. They might specify, for example, that files start with a particular style of block comment; that imports are ordered in a particular manner; that the first item in a class is the main program if the class has one; that the next items are field definitions and the field definitions are preceded by a block comment of a certain form and that public fields precede private ones; that any constructors follow the fields again preceded by a particular block comment; and so on. Examples of such standards include [6,9,14,19,21]. Even when the standards do not prescribe an order, as with the Google Java Style Guide [10], which notes that "what is important is that each class uses some logical order, which the maintainer could explain if asked".

What is in the various standards, however, is not particularly comprehensive or complete. Programmer's typically follow their own, more detailed ordering conventions [2]. For example, while the above guidelines provide general ordering information, they do not include all types of components (e.g. enumerations, factory methods, annotations). Moreover, they do not differentiate between orderings of components in an interface versus in a class, and they do not differentiate between the orderings and comment styles for inner classes versus outer classes. Both our own code and code retrieved from open source repositories shows that these orderings are not generally the same.

Programming environments benefit from understanding and being able to use program orderings. Today's environments provide a number of facilities that insert code, for example for automatically fixing errors or doing refactorings. If the environment does not do the insertion the way the programmer might, it can make more work for the programmer than it was designed to save. Programming environments typically include a simplified program ordering model. IntelliJ has the most complex one [11]. This model lets one define sections with before and after comments and define the order of element in that region using element type and modifiers. It includes the abilities to group getters and setters together and to group methods implementing a common interface together.

But even this complex model does not meet the needs of programmers. It does not differentiate between classes and interfaces or outer and inner classes as programmers do. It does not provide the semantic-based ordering options that programmers often use, for example, grouping a private method used only once with its caller. Moreover, setting up the model is a complex process and requires considerable interaction on the part of the programmer.

The goal or our research is to develop a model of program ordering that is flexible enough to order components in the same way that the programmer might. At the same time we want to be able to define this model automatically by learning it from the programmer's existing code base.

## II. OVERVIEW AND CONTRIBUTIONS

In order to determine and use program order, one needs to have a model that is flexible enough to reflect how programmers actually want to order code. The first part of our research involves developing such a model.

To do this we looked at a variety of style conventions and then looked at a large body of open source code as well as over one million lines of source in our own repository. We attempted to determine what types of conventions were used by the programmers in ordering the elements within these files, concentrating on determining the factors that might affect the ordering. Based on this informal study and previous work on understanding ordering [2,8], we developed a comprehensive model for file organization. This model is detailed in Section III.

The model is defined in two parts. The first part breaks the code file into regions each containing a set of components, a fixed prefix and suffix comment (which might be empty or just a set of blank lines), and a comment (or set of blank lines) that separates components in the region. Different sets of regions are defined for classes, interfaces, inner classes and inner interfaces. This is similar to the model used by IntelliJ except that it allows for more options in terms of selecting the set of components and differentiates the orderings based on their context. The second part of the model defines the order of the components within a region.

The model is designed to be used by a programming environment to determine where to best place a component, either a new one or an existing one if a file is being rearranged by the programming environment. The algorithm we use for this purpose is described in Section IV.

Because the model is relatively complex and we wanted to integrate it into an environment, we did not want to require the programmer to specify it manually. Thus the second part of our research is a learning framework that can build the above models based on a corpus of existing code that follows some set of conventions.

The learning framework, detailed in Section V, works in stages. It first learns the groupings of components into regions. It finds what the primary groupings are and the criteria used for groupings. Next it learns the ordering of these regions, merging regions where the orderings might not be significant. Next it learns the ordering of components within a region. The framework finishes by generating a machine-readable file that can be used by the programming environment to implement or to let the user edit the ordering.

While the framework is integrated into the Code Bubbles programming environment [17], we also evaluated it as a stand-alone system. We did two evaluations. The first ran the learner over a large number of open source projects and checked that the results seemed meaningful, doing checks to determine accuracy. The second study concentrated on a single, moderate-sized (more typical) project, randomly chose components from the project, and then used the model to reinsert those components to compare their actual position to the computed position. These studies and our conclusions based on them are detailed in Section VI.

The contributions of this work are:
- A comprehensive model for ordering program components that is flexible enough to reflect most of the wide variety of actual orderings programmers use including comments and spacing.
- A learning approach that can build this model from a code corpus.
- A validation that the model can be used to insert or reorder code in a programming environment.

III. THE ORDERING MODEL

Determining the correct location of a new component (method, field, inner class, etc.) in a source file requires having a model that lets one find the location. We wanted a model that would accommodate a wide range of programming styles and options.

Our model was implemented for Java. In Java, a component is represented by a declaration. This can be a class declaration, an interface declaration, an enumeration, a field declaration, a method declaration, an annotation type declaration, an annotation member declaration, or a static initializer. Declarations can be nested in that types (classes, interfaces, annotation types, and enumerations) can include other declarations. A file typically consists of a single type declaration. (For now we ignore the case where there are multiple top-level types in a file.) Our model provides a means for ordering the declarations inside a type. Much of the work should transfer to other languages. However additional constraints such as the need to place definitions before their uses, file scopes, and name spaces would have to be added and the properties considered might differ.

Our model views the contents of a type as an ordered sequence of *regions*. Each region consists of a set of components that share some set of properties, which we call a *category*. Each region also has an associated prefix comment, a comment that separates the components in a region, and an associated optional suffix comment. Each of these comments might be a specific one (e.g. a block comment indicating Private Fields), might be a generic one (a shaped block comment with different text in different cases), might just be a sequence of $k$ blank lines, or might be empty. A region with no components does not appear in the file. A region with components appears in the file as:

PREFIX COMP ( SEPARATOR COMP )* SUFFIX

where COMP represents the components in turn, PREFIX is the prefix comment, SEPARATOR is the separator comment, SUFFIX is the suffix comment, and * indicates zero or more repetitions of the parenthesized items to allow multiple components with an appropriate separator.

Component categories are defined based on the kind of component, on properties of the component, and on naming conventions. The model works by comparing components pairwise to determine whether one should precede the other or if there is no particular order between the two. Categories are then viewed as contiguous sets of components that have no determined order.

The properties used to determine categories are shown in Fig. 1. The properties with a *{From,To}* prefix indicate two separate properties based on the individual components and refer to the first or the second component being compared respectively. The last four sets of properties in the table reflect possible naming conventions of the project that might affect ordering. These are determined by regular expressions matching the element names. While these are predefined, they can be changed or augmented for a given project by providing a property definition file.

The overall model provides different orderings for different nesting types or *contexts*, in particular for classes, interfaces, inner classes and inner interfaces, with enumerations being treated as classes. This reflects the fact that pro-grammers sometimes treat these cases differently. For example, fields in an interface are actually constants and are often interspersed with abstract methods; the outer class might have block comments and multiple lines between methods, inner classes might not have comments and only have a single line between methods.

| Property | Description |
| --- | --- |
| NESTED | Indication of the nesting or context of the components, one of {CLASS, INTERFACE, INNER_CLASS, INNER_INTERFACE}. Enumerations are treated as classes here. |
| {From,To}TYPE | Type of the first or second component, one of CLASS, INTERFACE, ENUM, FIELD, METHOD, CONSTRUCTOR, INITIALIZER, ANNOTATION, or ANNOTATION_MEMBER |
| {From,To}PROTECT | Protection of the first or second component {PUBLIC, PROTECTED, PACKAGE, PRIVATE} |
| {From,To}STATIC | Flag indicating whether the first or second component is STATIC |
| {From,To}ACCESS | Flag indicating whether the first or second component is an access method, i.e. if its name matches the regular expression (get|is|set)[A-Z][A-Za-z0-9]*. |
| {From,To}FACTORY | Flag indicating whether the first or second component is a factory method, i.e. if its name matches the regular expression (new|create)[A-Z][A-Za-z0-9]*. |
| {From,To}OUTPUT | Flag indicating whether the first or second component is a getter method, i.e. if its name matches the regular expression (toString). |
| {From,To}MAIN | Flag indicating whether the first or second component is a main method, i.e. if its name matches the regular expression (main). |

Fig. 1. Properties used in determining the categories for a component based on comparing pairs of elements. The prefis {From,To} indicates two properties, one for the first element and one for the second.

| PROPERTY | DESCRIPTION |
| --- | --- |
| ALPHAORDER | Alphabetical ordering of the names {LSS,EQL,GTR} |
| CASEORDER | Case-insensitive alphabetical ordering of the names {LSS,EQL,GTR} |
| FIELDORDER | If both names start with one of {set, get, is, new} then the case-insensitive alphabetic ordering of the remainder of the name {LSS,EQL,GTR}, otherwise NA. |
| CALLORDER | One of <NONE,CALLS,CALLEDBY,BOTH> that describes whether the first method calls or is called by the second. NA if either of the components is not a method. |
| INTERFACEORDER | If both components implement an interface method, then one of {SAME,PRIOR,AFTER} to denote the order of the interface in the list of interfaces, otherwise NA |
| MOREPARAMS | If both components are methods with the same name, then T if the second has more parameters than the first and F otherwise. If either component is not a method or the names differ, then NA. |
| LENGTHORDER | Determines is the first component is significantly longer than the second. |
| {From,To}PROTECT | Protection of the first or second component {PUBLIC, PROTECTED, PACKAGE, PRIVATE}. |
| {From,To}STATIC | Flag indicating whether the first or second component is STATIC. |
| {From,To}FINAL | Flag indicating whether the first or second component is FINAL. |
| {From,To}ABSTRACT | Flag indicating whether the first or second component is ABSTRACT. |
| {From,To}CALLERS | Value indicating how many callers the first or second method has. Values include NA for non-methods, NONE, ONE, TWO, and MANY to indicate more than 2. |
| {From,To}GETTER | Flag indicating whether the first or second component is a getter method, i.e. if its name matches the regular expression (get|is)[A-Z][A-Za-z0-9]*. |
| {From,To}SETTER | Flag indicating whether the first or second component is setter method, i.e. if its name matches the regular expression (set)[A-Z][A-Za-z0-9]* |

Fig. 3. Properties used in determining the order of components within a region.

The model defines separate categories for each context. Each category consists of a set of one or more component types and a possibly empty set of properties which map a name to a set of values. A component is in a particular category if it has the same context as the category, if the component type is one of the component types of the category, and for each property in the set of properties, the value of that property for the component is in the associated set of values. Example of categories are shown in Fig. 2. In the figure, the first two categories are determined completely by the context and the element type. The third category is determined by the context, the type, and a set of properties. A category can also be defined as the union of two or more specific categories.

Each region has an optional ordering associated with it that determines how components are placed in the region; if the ordering is missing then new components are placed at the end of the region. The ordering can be based on the names of the components, on protection or other component properties, on calling properties of the two methods, on the number of parameters, or on the kind of method. The actual properties used and their descriptions are given in Fig. 3. The regular expressions for matching the element names for the last two property sets can be changed or augmented for a given project in the project-specific property definition file.

```
CATEGORY 1:
    CONTEXT: CLASS
    TYPE:    CLASS, INTERFACE, ANNOTATION,
             ANNOTATION_MEMER
    PREIFX:  5 blank lines
    SUFFIX:  1 blank line
    BETWEEN: 5 blank lines

CATEGORY 13:
    CONTEXT: CLASS
    TYPE:    FIELD
    PREFIX:  5 blank lines
    BETWEEN:
    SUFFIX:  1 blank line

CATEGORY 24:
    CONTEXT: CLASS
    TYPE:    METHOD
    PROPS:   ACCESS=T, PROTECT={PUBLIC}, STATIC=F
    PREFIX:  <block comment saying Access Methods>
    BETWEEN: 5 blank lines
    SUFFIX:  1 blank line
```

Fig. 2. Examples of categories used as part of the model as generated by the learner.

Orderings are represented in the model as a decision tree based on these properties. The decision tree compares two elements and provides one of three possible outcomes: the first should precede the second; the second should precede the first; or no ordering can be determined between the two.

The complete model is represented as an XML file which lists the different regions in the order they appear in their context, providing the selection criteria in terms of sets of categories, the prefix, suffix and between comments, and the ordering decision tree for each. It also includes the regular expression patterns that were used for name-based properties. The environment can read and store the model as well as let the programmer edit the model and save it again.

## IV. USING THE MODEL

Given this ordering model, the proper location and context for adding a new program element to an existing component can be found automatically. This involves translating comparisons between individual elements into a file location. Our algorithm for doing this is shown in Fig. 4.

This algorithm first (line 1) finds the ordering associated with the component into which the new component is being inserted. This is done by determining the nesting type associated with the component and choosing the corresponding ordering from the model.

Next, at line 3, it determines the region associated with the new program element using the function *findRegion*. This function goes through the regions listed in the ordering passed as the first argument, and checks, for each region, whether the new component passed as the second argument has a TYPE and property values that are consistent with the category descriptors for that region.

The loop between line 5 and line 17 attempts to determine elements (nested components) in C that should precede (PriorE) and follow (NextE) the element being inserted (NewE). It goes through the existing components E of C in order. For each component, it determines if it is in a prior region, the same region as NewE, or a subsequent region. If it is a prior region, then it sets PriorE to E since

Given a new program element NewE and a component C to add to:

```
1   Ord = the ordering associated with component C
2   NextE = PriorE = null
3   Rgn = findRegion(Ord,NewE)
4
5   FOR EACH program element E IN C IN ORDER :
6     Ergn = findRegion(Ord,E)
7     IF Ergn precedes Rgn in Ord THEN PriorE = E
8     ELSE IF Ergn == Rgn THEN
9       IF Rgn has ordering AND
10            E follows NewE in Rgn THEN
11        IF NextE == null THEN NextE = E
12      ELSE IF Rgn has ordering AND
13            E precedes NewE in Rgn THEN
14        PriorE = E
15        NextE = null
16      ELSE IF NextE == null THEN PriorE = E
17  END
18
19  Pos = Start of C
20  IF PriorE != null THEN
21    Pos = End of PriorE
22    NextE = the program element following PriorE in C
23  IF PriorE != null AND
24           findRegion(Ord,PriorE) == Rgn THEN
25    Add Between comment of Rgn in front of NewE
26  ELSE IF NextE != null AND f
27           indRegion(Ord,NextE) == Rgn THEN
28    Pos = start of NextE
29    Add Between comment of Rgn in back of NewE
30  ELSE
31    Add Prefix comment of Rgn in front of NewE
32    Add subbix comment of Rgn in back of NewE
33  END
34
35  Insert NewE at position pos
```

Fig. 4. Insertion Algorithm

NewE should come after the component. If it is a subsequent region and NextE is not set, then it sets NextE to E since in this case NewE should precede E. Finally, if E and NewE are in the same region, it looks at the ordering associated with the region for NewE and E. If NewE is known to follow E, then it makes PriorE = E and clears NextE. If NewE is known to precede E, then it sets NextE to E. If no ordering is known, and NextE is not set, then it sets PriorE to E, effectively moving the insertion toward the end of the region when no other ordering information is present.

Note that there may be multiple existing elements in the component that have an ordering relationship with the new component. The algorithm does not assume that the existing order is consistent with the ordering of the model. In this case, the algorithm is designed to prefer placing the new element towards the end.

Next, in line 19 through line 33, the algorithm determines what comments should be added to the new component based on the region associated with the prior and next components previously computed. It also determines the actual position of insertion from these components. Finally, the algorithm inserts the new component, with its comments, at the target position.

## V. LEARNING THE MODEL

Where today's programming environments provide ordering functionality, they require the user to manually define the ordering model. Especially where the model is relatively sophisticated as in IntelliJ, this can be a tedious process involving multiple dialogs. Our model is significantly more complex than the existing models. Defining the model manually would require much more interaction with the programmer. We wanted to minimize the amount of effort in using the model. As such we created a module to learn the ordering from an existing code base.

The learning process works in four stages: first finding the set of regions by determining what components go together; then determining the ordering of those regions; then determining the comment properties of regions; then determining the ordering of components in a region; and finally outputting a model.

*A. Finding the Regions*

We use WEKA [12,20] for learning. This is a standard, open-source platform that provides a variety of learning algorithms that use a consistent input format. The input to WEKA consists of a sequence of CSV-style lines each representing a known data point. The lines contain attributes of the data point followed by a result value. WEKA attempts to learn how to predict the result value based on the attributes.

The overall algorithm for learning the model is shown in Fig. 5. The main function, *BuildOrdering* constructs the ordering.

*BuildOrdering* starts by setting up a data file for learning. To determine what components are grouped together into regions, the learner starts by splitting each existing file in the project into components and characterizing these components (line 2 to line 6). This is done by adding a sample datum (WEKA CSV line) for each pair of subcomponents of either the top-level component or a subcomponent giving the properties of the two component as well as their order (*BEFORE*, *AFTER*). In addition, a sample datum is added for each component comparing it to itself with an order of *EQUAL*. This occurs in the recursive method *BuildModel* (line 21 to line 32) and the method *AddSample* (line 34 to line 40) that generates the data line. The attributes for each sample data point are those listed in Fig. 1.

The goal of the learning is to determine what orderings are actually used in a relatively consistent manner by the programmer(s). In particular, there are many cases where the ordering between elements in the code base is incidental and not particularly meaningful, and there are other cases where programmers do not strictly follow orderings. We need to represent these as *DONT-CARE*. Since we do not know a priori what elements fall in this category (other than an element compared to itself), we need to extract this after the fact from the learning output. These cases show up in the output as inconsistent orderings, i.e. an ordering where A precedes B and B precedes A, or in orderings that are not statistically significant given the overall corpus.

The algorithm next builds a decision tree that should be able to predict, given two subcomponents of a component, whether one follows or precedes the other (line 7). A decision tree lets us easily find cases where the generated ordering is inconsistent or statistically insignificant. We looked at the various decision tree approaches that are provided by WEKA. Some, such as HoeffdingTree and RandomForest generated trees that were too small, i.e. did not yield any meaningful orderings even when we knew ones were present. Others such as RandomTree generated trees that were too large, i.e. that included too many spurious orderings. The best methods were RepTree and J48 which tended to generate similar trees. From these we chose the J48 algorithm [15]. We looked at changing the various parameters associated with the J48 algorithm and eventually settled on the default parameters for generating the most appropriate tree. These decisions were made by running the different learning algorithms (or J48 with different parameters) on a corpus where we knew the approximate ordering.

The resultant decision tree is then cleaned up using the method *CleanUp* (line 42 to line 53) The clean-up process is designed to add *DONT-CARE* decisions to the tree by replacing decisions with EQUAL that are not statistically significant or that are logically inconsistent. (*EQUAL* is then interpreted as *DONT-CARE*.) The method first (line 43 to line 46) checks that the number of samples that follow the decision is statistically significant (more than one standard deviation unit from the mean and at least five instances). If it is not, it replaces the decision with *EQUAL*. Next the method checks the tree for consistency (line 47 to line 53). If the tree indicates that component A should precede component B, it should also indicate that component B should come after component A. For each leaf, it checks the prediction using the tree where all the *From_* attributes are swapped with the *To_* attributes. If the result is not the opposite of the original prediction, then the tree is modified to say the two cases are *EQUAL*.

The next step is to construct all possible relevant sets that can define potential regions (line 9). The method *BuildInitialRegions* (line 102 to line 109) does this by considering each possible context (CLASS, INTERFACE, INNER_CLASS, or INNER_INTERFACE) and each possible symbol type. For each pair of these, it looks at the decision tree and finds the set of properties that are used for this pair. Then it builds an initial region set for each possible setting of each of these properties. This typically yields two to three hundred potential region sets. These are cleaned up and merged later in the process.

Given the refined decision tree and the initial region sets, the framework next goes through each file in the project again to find the sets that define the regions where each set is characterized by a sample component given its context and properties using the method *OrderRegion* (line 55 to line 71). For each component in the file, it determines the set associated with that component using the method *FindRegion* (line 73 to line 78). This finds the region that is consistent with the given element and updates the number of elements associated with the region. Since the set of initial region sets covers all possibilities by construction, this is guaranteed to find a region.

During this pass, the comments associated with the regions and the ordering of the regions is determined. The ordering of regions is done by building a weighted graph where the nodes are the regions and the arcs between these regions are weighted by the number of times the source region precedes the target. For each file, the sequence of region sets in the file for each context is determined by taking each component, determining its region set, and then noting when the region sets change (line 62 to line 65) by

```
1   Function BuildOrdering()
2     FOREACH File F
3       FOREACH top level component C IN F
4         IF C is an interface THEN
5           BuildModel(C,INTERFACE)
6         ELSE BuildModel(C,CLASS)
7     Tree = BUILD a J48 Decision tree using WEKA
8     CleanUp(Tree)
9     BuildInitialRegions(Tree)
10    Graph = new empty graph
11    FOREACH File F
12      FOREACH top level component C in F
13        IF C is an interface THEN
14          OrderRegion(C,INTERFACE,Graph,Tree)
15        ELSE OrderRegion(C,CLASS,Graph,Tree)
16    MergeEmptyRegions()
17    SortGraph = CleanGraph(Graph)
18    Order = TopSort(SortGraph,Graph)
19    return Order
20
21  Function BuildModel(C,Ctx)
22    FOREACH Program Element E1 IN C
23      IF E1 is an interface THEN
24        BuildModel(E1,INNER_INTERFACE)
25      ELSE IF E1 is a class/enum THEN
26        BuildModel(E1,INNER_CLASS)
27      AddSample(Ctx,E1,E1,EQUAL)
28      FOREACH Program Element E2 in C
29        IF E1 precedes E2 in C THEN
30          AddSample(Ctx,E1,E2,BEFORE)
31        ELSE IF E1!= E2 THEN
32          AddSample(Ctx,E1,E2,AFTER)
33
34  Function AddSample(Ctx,E1,E2,R)
35    S = new Sample data point for WEKA
36    Add Ctx to S
37    Add Properties of E1 to S with prefix From_
38    Add Properties of E2 to S with prefix To_
39    Add Result R to S
40    Output sample to data file
41
42  Function Cleanup(Tree)
43    FORALL nodes N of Tree
44      Look at # sample agree/disagree with
            Result(N)
45      IF the # is not statistically significant
          THEN
46        Result(N) = EQUAL
47    FORALL nodes N of Tree
48      R = Result(N)
49      Compute N1 = Node of tree when FROM and TO
50          of N are swapped
51      R1 = Result(N1)
52      If R and R1 are not opposites THEN
53        Result(N) = EQUAL
54
55  Function OrderRegion(C,Ctx,G,T)
56    PrevGroup = null
57    FOREACH Program Element E of C
58      IF E is an interface THEN
59        OrderRegion(E,INNER_INTERFACE,G,T)
60      ELSE if E is a class/enum THEN
61        OrderRegion(E,INNER_CLASS,G,T)
62      Group = FindRegion(Ctx,E)
63      IF Group != PrevGroup THEN
64        IF PrevGroup != null THEN
65        Add PrevGroup->Group to G
66        IF PrevGroup != null THEN
67          Record Comment for end of PrevGroup
68        Record Comment for start of Group
69        PrevGroup = Group
70      ELSE
71        Record Comment for between of Group
72
73  Function FindRegion(Ctx,E)
74    FOREACH Region R associated With Ctx
75      IF E is consistent with R THEN
76        Add E to R
77        Return R
78    RETURN R1
79
80  Function CleanGraph(Graph)
81    NewGraph = new empty Graph
82    FOREACH Arc A->B in Graph
83      If Significant(A,B,Graph) THEN
84        Add A->B to NewGraph
85    RETURN NewGraph
86
87  Function TopSort(SortGraph,Graph)
88    Do a topological sort of Graph
89    When adding N to Output with P the prior output
90      IF P == null OR Significant(N,P,Graph) THEN
91        add N to Output as new Group
92      ELSE Add N to previous group
93
94  Function Significant(A,B,Graph)
95    W1 = Weight(A->B) in graph
96    W2 = Weight(B->A) in graph
97    IF W1 is statistical > W2 THEN
98      RETURN True
99    ELSE
100     RETURN False
101
102 Function BuildInitialRegions(Tree)
103   ForEach N In NestType.values()
104     ForEach ST in SymbolType.values()
105       Set Props = Properties in Tree for <N,ST>
106       Construct new Region for each possible
            value
107         of a property in Props
108     NEXT
109   NEXT
110
111 Function MergeEmptyRegions()
112   FOREACH R in Regions with no Elements
113     FOREACH MR in Regions
114       IF MR has the same context as R
115         AND MR has associated elements
116         AND MR has only one property P different
117       Merge R with MR using property P
118       BREAK
119     NEXT
120   NEXT
```

Fig. 5. Learning the Regions

adding the transition to the graph. At the same time, the method records the comments that exist in the code either for starting a region, between elements in a region, or at the end of the region.

Once this is done, the framework eliminates any of the initial region sets that are empty. Empty region sets can occur for two reasons. First, some property settings are inconsistent with one another. For example, the name patterns for ACCESS, FACTORY, OUTPUT, and MAIN are all mutually exclusive. Also, the toString method (OUTPUT), when overriding the default, is always public and non-static. Second, there are valid cases that just might not arise in the sample code. For example, creating nested classes inside inner classes is rare.

For each empty region set, the framework attempts to find a non-empty region set that differs from the empty set in only one property (line 16) using the method *MergeEmptyRegions* (line 111 to line 120). If such a set can be found, then the two sets are merged by adding the alternative value of the property to the non-empty set. If all possible values of the property are included, for example both T and F values are included for STATIC, then the check for that property is eliminated in the merged set.

The next step is to clean up the graph generated above so that there is an arc from region set A to region set B if and only if A generally precedes B (line 17). This is done using the method *CleanGraph* (line 80 to line 85). Significance is determined in method *Significant* (line 94 to line 100) by checking if A precedes B significantly more often than B precedes A (more than one standard deviation unit) and A precedes B a minimum number (currently four) times.

Once the graph is built, it is converted into a linear ordering using a modified topological sort that always produces an ordering (line 18). This is done in method *TopSort* (line 87 to line 92). The order produced by the sort might not be definitive. There might be multiple region sets valid at a given point (or a cycle of such sets) or the ordering might be a side effect of the sort rather than an actual ordering. These are both checked by checking for each region in the resultant ordering if the graph indicates it is significantly different from its predecessor. If not, then the region is merged with the previous forming a union region. The result of this is, for each context, a sequence of regions, some of which are unions.

### B. Learning Comment Properties

The comment properties of each region are determined from the information gathered while finding the regions for each component in method *OrderRegion*. If the component is in the same region as the previous component, the text between the components (comments and blank lines) is recorded as a potential separator for the region. If the component is in a different region, the text between it and the prior component (or the start of the context) is split if there are multiple block comments and is recorded as a potential suffix for the prior region and prefix for the new region.

The prefix, suffix, and separator for each region are then determined by looking at the set of text regions that were gathered for the region in this way. The framework counts the number of instances of each text block and finds the mode instance (the one that occurs most often). If this occurs at least four times and at least 25% of the time, it is assumed to be the appropriate result. Otherwise, the framework attempts to simplify comments and then recheck. The first simplification replaces any text in a comment with blanks, trimming lines as appropriate. This handles the case where the comment has a standard form but different contents. If this simplification does not yield an acceptable result, then a second simplification is tried that not only removes all text but also removes duplicate lines once the text is removed. This will simplify all JavaDoc comments into a standard form regardless of their length. If this simplification yields an acceptable result, it is used. Otherwise, a set of $k$ blank lines, where $k$ is the median number of lines between components, is used as the comment.

### C. Learning Ordering within a Region

The ordering properties of each region are determined by building a second J48 decision tree using WEKA. For each pair of components in the same region, a sample datum is generated using the properties listed in Fig. 3 along with a property INDEX which indicates the region identifier. The properties describe the context, the properties of the two components, various comparators (alphabetic, parameter counts, alphabetic ignoring prefix, alphabetic ignoring case, length), and the resultant order (*BEFORE*, *AFTER*, or *EQUAL*). The resultant decision tree is cleaned up using the same techniques as the original tree, i.e. ensuring consistency and significance to effectively add a *DONT-CARE* output (*EQUAL*) to the tree.

### D. Outputting the Model

The final step is to create and output the resultant model from this information. The region sets built are already specialized for a particular context and include the relevant component types and other properties.

Each region also has an associated ordering decision tree. This is built for the region by taking the overall ordering decision tree and specializing it for this particular region. The region comes into play if the tree includes a branch using the property INDEX or where there are properties (e.g. STATIC or PROTECT) that are used for both trees. Each output region reflects a set of region ids. A new decision tree is built from the original by replacing branches that involve the region id with the branch that would result from using the actual region id. If the result is not consistent, then the branch is replaced with an *EQUAL* (representing *DONT-CARE*).

The overall model is output in the XML format that is accepted by the Code Bubbles programming environment.

## VI. EXPERIENCE AND EVALUATION

We evaluated the approach to modeling in two ways. First, we built models for over 500 different open source projects to ensure the system ran within reasonable time limits and spot checked a number of the resultant models, especially ones for which there was a standards document. Second, we built a model of program ordering for a known corpus and then checked how well the insertion algorithm worked based on that model.

### A. Building Program Models

We used the DARPA Muse repository and created a script to go through the different packages. For each package we counted the number of non-test Java source files (files that did not use JUnit annotations or that were not in a test directory hierarchy) and then, if there were more than 200 such files, we built the corresponding model using the above learning algorithm. We did spot checks to ensure that the resultant models were logical. Then we looked for coding standards on the web and where possible, compared those standards to the models that were built.

One standard we found was for the Apache ACE project (https://ace.apache.org/docs/coding-standards.html). This standard specifies the order as static variables grouped by functionality, instance variables, constructors, methods (grouped by functionality rather than scope or accessibility), and finally inner classes. The deduced ordering was somewhat different: a) Inner interfaces, classes, and enumerations; b) Static fields and static initializations; c) Instance fields; d) Constructors; e) Public methods; f) Package methods; g) Protected methods; h) Private methods; and h) Output methods. The major differences were that we detected an ordering within the methods that was not necessarily specified but was followed by the programmers, and we detected that inner classes (and interfaces and enumerations) were at the start of the file rather than the bottom. We checked the actual sources and, for the most part, this is indeed the case.

The Apache Felix project has similar standards. Here the deduced order was: a) Enumerations and static fields; b) Instance fields; c) Static initializers; d) Constructors; e) Non-private methods; f) Non-public methods; and g) Inner classes. This is a closer match to the given standards.

The time required to build program models seems reasonable. For a system with 18,000 lines of source (50 files), the time was 15 seconds; for the moderate sized project used in the detail experiments below, which was 88,000 lines of source in 235 files, the time was 91 seconds; for a large project (306,000 lines, 732 files) the time was 9.5 minutes.

*B. Analyzing Model-Based Insertions*

To understand the utility of the models and the insertion algorithm based on them, we did a second experiment. Here we took an open source project we knew well from GitHub (https://github.com/StevenReiss/s6) and built the model based on its non-test source files. The project included about 88,000 lines of source in 235 files. We excluded ten randomly chosen files from this set to use for later testing. In each of these ten files, we randomly selected four components (one only had three) to be inserted. We removed these components and then used the model and algorithm to reinsert it into the file. The use of existing components in files that were not part of the learning process lets us test the learning output and insertion algorithm against what is effectively the ground truth, what the programmers actually did.

The learning algorithm produced a logical ordering that pretty much matched our analysis of the files. The deduced ordering for classes was:
- Interfaces, non-static fields, enumerations, main method
- Static fields
- Initializers
- Constructors
- Static factory methods
- Non-static factory and access methods
- Other methods
- Output methods, static access methods
- Other static methods, inner classes

The deduced order for inner classes was:
- Fields, static methods
- Constructors, initializers, inner type definitions
- Factory methods, access methods
- Other methods

- Output methods

The deduced order for interfaces put inner interfaces last and everything else before that. There was no deduced order for inner interfaces (i.e. their components can occur in any order).

The differences between the deduced order and the expected one were of two types. First, some of the categories that include multiple types probably should have separated those types. More recent files put the main method first, but this was not always true over the evolution of the project, and hence was not recognized by the learner. Initializers typically come before constructors in inner class, but this occurs so infrequently that it is not significant. The second type of difference introduced unexpected categories. For example, static access methods were not expected to be separate from access methods or put at the end of the file. This case arose because static access methods were uncommon in the code base and tended to occur together in files that did not have non-static methods.

Fig. 6 shows the results of the insertions. The first column gives the file name; the second is the type of item being inserted; and the third is the name relative to the outer class. Qualified names in this column indicate inner classes or interfaces. The fourth column is the difference in terms of the number of characters between the original and computed location. This is generally going to be positive since the algorithm has a bias toward placing new elements towards the end of the file. The cases where the delta is zero (16/39) indicate that the framework placed the component in the exact place it had been. The one case with a large error (*SolutionAdder*) represents an inner class that was originally in the middle of the file and was placed at the end.

The next column indicates whether the element was placed at the end of its region. With our learner and the repositories this is generally going to be the case since only a minimal ordering specification was learned. What was learned was to place callers before the methods they call; to place methods with the same name in order based on the number of parameters; and to group methods from the same interface. The random sample of elements that were chosen had no such examples. The fact that there are a significant number of different categories make this somewhat less of a problem as noted by the number of elements that were placed correctly.

*C. Limitations*

The various experiments showed that overall the method works, although there are a number of caveats.

First, the approach is sensitive to the size and nature of the corpus that is used for training. The above experiment, with about 88,000 lines of source, is a moderate sized repository. It was built by multiple programmers over several years, and the orderings used by the programmers tended to change slightly over time (such as where the main method was placed). Building it over a smaller repository (18,000 lines), tended to yield more errors because unusual pairings of elements do not occur frequently enough. For example, public methods in inner classes were misplaced. Note that this occurred in the moderate sized repository as well with initializers and constructors in inner classes. Another problem with small repositories is that one unusual file can skew the results. In the repository we used, there were a small number of inner classes with a large number of public overridden methods that caused the issues with public methods. Using an even larger repository has more problems with multiple programmers and changes over time. It also tends to produce more categories with some of the rankings being spurious. This could probably be fixed by making the notion of significance stricter.

| File | Type | Name | Delta | Region End |
|---|---|---|---|---|
| S6Factory.java | METHOD | createCheckRequest | 355 | true |
| S6Factory.java | METHOD | createKeySearch | 192 | true |
| S6Factory.java | METHOD | createSolutionSet | 121 | true |
| S6Factory.java | METHOD | createLanguage | 51 | true |
| ContextEclipse.java | FIELD | current_workspace | 0 | true |
| ContextEclipse.java | CONSTRUCTOR | <init> | 0 | true |
| ContextEclipse.java | METHOD | setWorkspace | 1650 | true |
| ContextEclipse.java | METHOD | getClassPath | 0 | true |
| EnginePool.java | FIELD | num_io | 109 | true |
| EnginePool.java | CONSTRUCTOR | <init> | 0 | true |
| EnginePool.java | CLASS | Worker | 0 | true |
| EnginePool.java | FIELD | Worker.worker_index | 0 | true |
| KeySearchSource.java | FIELD | search_index | 0 | true |
| KeySearchSource.java | CONSTRUCTOR | <init> | 0 | true |
| KeySearchSource.java | METHOD | getProjectId | 157 | true |
| KeySearchSource.java | METHOD | getDisplayName | 78 | true |
| TransformJava.java | CLASS | SolutionAdder | 12919 | true |
| TransformJava.java | FIELD | SolutionAdder.solution_set | 76 | true |
| TransformJava.java | FIELD | TreeCopy.new_base | 0 | true |
| TransformJava.java | FIELD | JavaMemo.use_position | 28 | true |
| TransformFixEnum.java | CLASS | EnumNameMapper | 0 | true |
| TransformFixEnum.java | FIELD | EnumNameMapper.fix_names | 0 | true |
| TransformFixEnum.java | METHOD | EnumNameMapper.getSpecificsName | 0 | true |
| TransformFixEnum.java | METHOD | EnumNameMapper.rewriteTree | 0 | true |
| RunnerTest.java | FIELD | AddPlayerFrame.aliasCheckbox | 135 | true |
| RunnerTest.java | FIELD | AddPlayerFrame.passCheckbox | 36 | true |
| RunnerTest.java | METHOD | AddPlayerFrame.getAccountPanel | 1348 | true |
| RunnerTest.java | METHOD | AddPlayerFrame.addPlayer | 0 | true |
| RunnerPencilHierData.java | FIELD | child_elements | 0 | true |
| RunnerPencilHierData.java | METHOD | getTypes | 1499 | true |
| RunnerPencilHierData.java | METHOD | getRightAnchor | 1295 | true |
| RunnerPencilHierData.java | METHOD | setTopAnchor | 751 | true |
| SearchWordConstants.java | FIELD | WORD_LIST_FILE | 1125 | true |
| SearchWordConstants.java | ENUM | WordOptions | 482 | true |
| SearchWordConstants.java | ENUM | TermOptions | 0 | true |

Fig. 6. Results of insertion experiment. The first column is the file name; the second is the type of item being inserted; the third is the name relative to the outer class; the fourth is the difference in terms the number of characters between the original and computed location; the final indicates that whether the component was placed in the end of the region or not.

A second problem is that programmers (and standards) attempt to order methods by their functionality or semantics. We attempted to capture some of this in the CALLORDER and CALLERS properties and by the INTERFACEORDER property, but these are only rough approximations. Finding other means of determining functionality would be helpful. This becomes more of a problem when adding new code since the resultant method body is empty and determining functionality would be even more difficult. Other orderings proposed in the literature [2] such as grouping fields with common data types together, and grouping overridden methods together could be accommodated with additional fields. Other orderings, such as grouping entities together that are inserted together, grouping by frequency of usage, or placing entities in order of execution are difficult to determine a priori from an existing code base.

A third problem is in categorizing methods, for example, determining which methods are getters and which are setters. Our current approach looks only at the names. However, programmers often consider methods to be getters or setters based on properties other than the names. This can work both ways. A method might have the form getXXX but actually do significant computation and effectively not be a getter. Similarly, a simple method such as length() might be considered as a getter even though its name does not start with *get* or *is*. Method names starting with "new" might be considered either factory methods or getters depending on their semantics. Again, to work for new methods, an alternative categorization would not have access to the method body, only the name.

A fourth potential problem is that the current approach might not consider all properties used by all programmers in determining order. For example, one possible approach to grouping overridden methods together would be to include an OVERRIDES property which indicates whether the method overrides a method in a super class or interface. This however, can be difficult to determine since the learner might not have access to all libraries and hence might not be able to detect which methods were overridden. (It could, however, check for the @Override annotation relatively easily; this however, might not be used consistently, especially in older Java systems.)

One thing we noted in all the examples, is that the learner rarely finds significant orderings among the elements in a grouping other than based on the CALLORDER and MOREPARAMS properties. This seems reflective of how the files are actually ordered in that methods were either grouped by function or by when they were inserted.

Because of these problems, the orderings that programmers actually want might not be completely reflected in the ordering that is learned. Because orderings change over time and between programmers, because programmers are not always strict about their use of ordering, and because it can be difficult to find an ideally sized repository, an alternative is to use the learned ordering as a first approximation and then to let the programmer edit the ordering to correct any perceived errors or to add additional constraints. Our approach allows this to occur, either by having the programmer directly edit the XML output or, more likely, by having the environment provide an appropriate editing interface. In any case, editing an ordering should be easier than attempting to construct one from scratch.

### D. Threats to Validity

While the experiments showed the learning works and the insertion algorithm behaves reasonably, there are several threats to assuming this is going to be true in all cases.

First, the code used in the experiments might not be typical. Other code might have a stronger or weaker ordering model imposed by the programmer or the coding standards. Anomalies in the corpora might adversely affect the results (as the public methods did for inner classes in the smaller example).

For the large scale model building experiments, we were not familiar with the actual code and could only perform spot checks as to the validity of the generated models. It is possible that we did not check models that were inherently different from what the programmer intended.

For the detailed experiments, we used code we knew relatively well and hence could be fairly certain that the properties used in ordering were included in the set of properties used to build the model. It is possible that to build accurate models for other systems, additional properties would be required. Note that it is fairly simple to add properties to the system and then build the corresponding models.

The detailed experiments looked at reinserting pre-existing methods into a code file. If one were to insert new methods, where the method bodies were not yet defined or where certain properties such as protection might not be accurate, the model might not do as well.

We also used a predefined set of naming conventions that seemed relatively standard and matched the code in the repositories we tested. It is possible that other repositories would use other conventions and that these would have to be changed to get appropriate results. Our system make this easy to do.

## VII. RELATED WORK

Machine learning has been applied to learning coding conventions. Corbo, et al. [7] demonstrated that one could learn the Eclipse formatting conventions from existing code. Reiss [16], and more recently Allamanis et al. [1], show that one can learn general formatting conventions as well as naming and other stylistic conventions from existing code with reasonably high accuracy. These systems are capable of changing the names as well as updating the styles globally. Reiss also included limited support for ordering, primarily deducing the ordering of import declarations, but did not build a general ordering model for components. There has also been general work on learning orderings [3,5].

The use of program ordering has been studied both in itself [2] and in terms of how it affects program understandability and maintenance [8]. The effectiveness of using coding standards, including program ordering, has also been studied in various ways, for example looking at how programmers search for code [13], the effect on errors [4], and the use of coding standards as metrics [18].

## VIII. CONCLUSIONS

This work shows that it is possible to derive relatively complex and sophisticated models for how components are ordered within a file. The overall framework includes the set of properties that may be relevant to component ordering. The learning model takes into account the fact that programs do not perfectly reflect the ordering desired by the programmer. It tries to determine what aspects of the corpus are actually significant in determining the ordering. The model also distinguishes between finding appropriate groups of elements and learning the ordering within groups.

The learning package, the example generation code and scripts, and the actual data used in the experiments are available from the authors on request.